\documentclass[12pt,thmsa]{article}
\usepackage{cite}
\usepackage{graphicx}

\input{tcilatex}

\begin{document}

\title{Alternative Fourier--series approach to nonlinear oscillations}
\author{Francisco M. Fern\'{a}ndez \\
INIFTA (UNLP, CCT La Plata--CONICET), Divisi\'{o}n Qu\'{i}mica Te\'{o}rica,\\
Diag. 113 y 64 (S/N), Sucursal 4, Casilla de Correo 16,\\
1900 La Plata, Argentina \\
e--mail: fernande@quimica.unlp.edu.ar}
\date{}
\maketitle

\begin{abstract}
We propose an alternative approach that avoids the nonlinear equations for
the Fourier coefficients that appear in the method of harmonic balance. We
apply it to two simple illustrative examples.
\end{abstract}

\section{Introduction}

\label{sec:intro}

Since one--dimensional nonlinear oscillators are treated as simple
pedagogical examples in many standard textbooks\cite{N73, BO78, NM79, N81,
S94} one is tempted to believe that they are trivial problems. However, from
time to time there appears to be a renewed interest in them as shown by a
recent unbelievable article\cite{RH09}. The equations of motion for the
simplest nonlinear oscillators are suitable for illustrating the application
of several approximate methods that include various forms of perturbation
theory, variational approaches, Fourier expansions, etc.\cite{N73, BO78,
NM79, N81, S94}.

The method of harmonic balance\cite{NM79}, which is based on the Fourier
expansion of the solution of a periodic nonlinear oscillator, yields
accurate results but requires the solution to a rather complicated system of
nonlinear equations for the expansion coefficients. It is not that difficult
to select the physical solution of a nonlinear equation in one variable, but
the problem becomes increasingly complicated as the number of equations and
variables increases (specially if there are many unwanted solutions besides
the physical one). The purpose of this communication is to propose a simpler
approach that avoids the nonlinear system of equations for the Fourier
coefficients. It is based on a systematic generalization of the approach
proposed by Ren and He\cite{RH09} in a paper plagued by errors and
misconceptions\cite{F09e}. In Sec.~\ref{sec:method} we develop the method
and in Sec.~\ref{sec:examples} we apply it to two simple but nontrivial
illustrative examples. Finally, in Sec.~\ref{sec:conclusions} we discuss the
advantages and limitations of the approach.

\section{The method}

\label{sec:method}

Consider the dimensionless equation of motion for a nonlinear oscillator
\begin{equation}
u^{\prime \prime }(t)+f[u(t)]=0,\,u(0)=A,\,u^{\prime }(0)=0
\label{eq:eq_mot}
\end{equation}
where $u(t)$ is the displacement from equilibrium and the prime indicates
differentiation with respect to the independent variable $t$. For simplicity
we assume that the force $-f(u)$ is an odd function of the displacement: $%
f(-u)=-f(u)$ and that the motion is bounded with a period $T=2\pi /\omega $,
where $\omega $ is the angular frequency. The potential--energy function $%
V(u)$ ($f(u)=dV(u)/du$) is, consequently, an even function of the
displacement ($V(-u)=V(u)$). Such a symmetry and the initial conditions
determine that $u(T/4)=0$ and therefore we can restrict our calculations to
the interval $0\leq t\leq T/4$ where $A\geq u(t)\geq 0$. Obviously, if we
know $u(t)$ in this interval then we know it for all values of $t$.

If we try to solve the equation of motion (\ref{eq:eq_mot}) by means of a
truncated Fourier--series expansion
\begin{equation}
u^{[N]}(t)=\sum_{j=1}^{N}A_{j}\cos [(2j-1)\omega t]  \label{eq:Fourier}
\end{equation}
and the method of harmonic balance\cite{NM79} then we are led to a system of
nonlinear equations for the expansion coefficients $A_{j}$. In what follows
we propose a simple strategy that avoids such a problem. Notice that the
ansatz (\ref{eq:Fourier}) satisfies $u^{[N]}(T/4)=0$.

If we define $u^{(2n)}(t)=d^{2n}u(t)/dt^{2n}$, $u^{(0)}(t)=u(t)$ then we can
write the set of equations $u^{(2n)}(0)=G_{n}(A),\,n=0,1,\ldots $, where $%
G_{0}(A)=A$, $G_{1}(A)=-f(A)$, $G_{2}(A)=f(A)f^{\prime }(A)$, etc.
Substitution of Eq.~(\ref{eq:Fourier}) into the left--hand side of these
equations for $n=0,1,\ldots ,N-1$ leads to the \textit{linear} system of
equations
\begin{equation}
\sum_{j=1}^{N}(2j-1)^{2n}\omega ^{2n}A_{j}=G_{n}(A),\,n=0,1,\ldots ,N-1
\label{eq:system}
\end{equation}
that gives us the coefficients $A_{j}$ in terms of the amplitude $A$. The
matrix of this system of equations is the transpose of a Vandermonde one and
therefore its determinant is nonzero.

If we now integrate the equation of motion (\ref{eq:eq_mot}) twice and take
into account the initial conditions we have
\begin{equation}
u(t)=A-\int_{0}^{t}\int_{0}^{t^{\prime }}f(u(t^{\prime \prime
}))\,dt^{\prime \prime }\,dt^{\prime }  \label{eq:u(t)}
\end{equation}
that we can apply iteratively. Here we carry out only one iteration
\begin{equation}
\tilde{u}^{[N]}(t)=A-\int_{0}^{t}\int_{0}^{t^{\prime }}f(u^{[N]}(t^{\prime
\prime }))\,dt^{\prime \prime }\,dt^{\prime }  \label{eq:u(t)[N]}
\end{equation}
and obtain the period from the resulting approximate solution $\tilde{u}%
^{[N]}(t)$ and the condition derived above:
\begin{equation}
\tilde{u}^{[N]}(T/4)=0  \label{eq:u(T/4)=0}
\end{equation}

For $N=1$ we simply have $u^{[1]}(t)=A_{1}\cos (\omega t)$ and $A_{1}=A$
from the only equation in (\ref{eq:system}). The system of equations (\ref
{eq:system}) with $N=2$ yields
\begin{eqnarray}
A_{1} &=&\frac{9A\omega ^{2}-f(A)}{8\omega ^{2}}  \nonumber \\
A_{2} &=&\frac{f(A)-A\omega ^{2}}{8\omega ^{2}}  \label{eq:A1,A2}
\end{eqnarray}
These two approaches of first ($N=1)$ and second ($N=2$) order are
sufficiently accurate as suggested by the examples below.

If the force is not an odd function of the displacement then we choose the
ansatz $u^{[N]}(t)=A_{0}+A_{1}\cos (\omega t)+A_{2}\cos (2\omega t)+\ldots
+A_{N}\cos (N\omega t)$, which should satisfy $u^{[N]}(0)=A$ and $%
u^{[N]}(T/2)=B$, and resort to the independent condition $\tilde{u}%
^{[N]}(T/2)=B$, where $B$ is the other turning point.

\section{Examples}

\label{sec:examples}

Our first example is one of the most widely studied nonlinear oscillators:
the Duffing equation given by the function\cite{N73, BO78, NM79, N81, S94}
\begin{equation}
f(u)=u+\epsilon u^{3}  \label{eq:Duffing_f(u)}
\end{equation}
It is not difficult to prove that $\rho =\epsilon A^{2}$ is the relevant
parameter of the model. That is to say, $u(t)/A$ and other properties of the
system, like, for example, the period, depend on $A$ and $\epsilon $ only
through $\rho $ and not separately or in other combinations.

In the first approximation we derive the period\cite{RH09}
\begin{equation}
T^{[1]}(\rho )=\frac{6\pi }{\sqrt{7\rho +9}}  \label{eq:Duffing_T[1]}
\end{equation}
In order to verify its accuracy for large values of $\rho $ we recall that
the exact period satisfies
\begin{equation}
\lim_{\rho \rightarrow \infty }\sqrt{\rho }T(\rho )=T_{\infty }=7.416298709
\label{eq:Duffing_T_inf}
\end{equation}
The approximate expression (\ref{eq:Duffing_T[1]}) yields the reasonable
estimate
\begin{equation}
\lim_{\rho \rightarrow \infty }\sqrt{\rho }T^{[1]}(\rho )=T_{\infty
}^{[1]}\approx 7.12  \label{eq:Duffing_T[1]_inf}
\end{equation}
For small values of $\rho $ we have
\begin{equation}
T(\rho )=2\pi -\frac{3\pi \rho }{4}+\frac{57\pi \rho ^{2}}{128}+\ldots
\label{eq:Duff_T_Taylor}
\end{equation}
and
\begin{equation}
T^{[1]}=2\pi -\frac{7\pi \rho }{9}+\ldots  \label{eq:Duff_T[1]_Taylor}
\end{equation}
We appreciate that $T^{[1]}(\rho )$ does only give us the leading term of
the small--$\rho $ series.

For $N=2$ we easily obtain
\begin{eqnarray}
&&125T^{8}\rho \left( \rho +1\right) ^{3}-7656\pi ^{2}T^{6}\rho \left( \rho
+1\right) ^{2}+120\pi ^{4}T^{4}\left( \rho +1\right) \left( 1607\rho
+735\right)  \nonumber \\
&&-64\pi ^{6}T^{2}\left( 48851\rho +55125\right) +12700800\pi ^{8}=0
\label{eq:T[2]}
\end{eqnarray}
The approximate period is given by the only real root $T^{[2]}(\rho )$ that
satisfies $T^{[2]}(0)=2\pi $. It is clear that we have not avoided the
nonlinear nature of the problem completely but we have reduced it to a an
equation with just one variable which simplifies the task of choosing the
physical solution.

Without difficulty we derive the following expression for the approximation
of second order to $T_{\infty }$
\begin{equation}
125T_{\infty }^{8}-7656\pi ^{2}T_{\infty }^{6}+192840\pi ^{4}T_{\infty
}^{4}-3126464\pi ^{6}T_{\infty }^{2}+12700800\pi ^{8}=0  \label{eq:T[2]_inf}
\end{equation}
from which we obtain $T_{\infty }^{[2]}\approx 7.44$ that is in fact more
accurate than $T_{\infty }^{[1]}$.

The Taylor expansion of $T^{[2]}(\rho )$ about $\rho =0$
\begin{equation}
T^{[2]}(\rho )=2\pi -\frac{3\pi \rho }{4}+\frac{1431\pi }{3200}+\ldots
\label{eq:Duff_T[2]_Taylor}
\end{equation}
exhibits the first two terms of the exact small--$\rho $ series.

\begin{figure}[]
\begin{center}
\includegraphics[width=9cm]{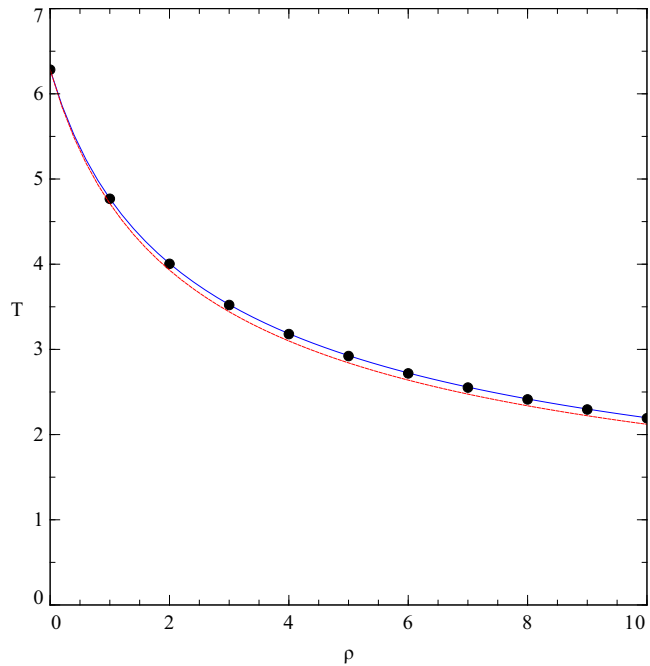}
\end{center}
\caption{Period for the Duffing oscillator. Circles, dashed and solid lines
stand for exact, first and second approximations, respectively}
\label{fig:Duff}
\end{figure}

Fig. \ref{fig:Duff} shows the exact period and the two approximations just
derived for small and moderate values of $\rho $. We appreciate that $%
T^{[2]}(\rho )$ is more accurate than $T^{[1]}(\rho )$ for all values of $%
\rho $ as expected from the analytic results derived above. It is
encouraging that the accuracy of the method increases notoriously with the
order and that just the lowest two approximations are in good agreement with
the exact result.

Our second example is given by
\begin{equation}
f(u)=u+\epsilon |u|u  \label{eqDO_f(u)}
\end{equation}
and we realize that the relevant parameter is $\rho =\epsilon A$. This
problem is as simple as the preceding one and we easily obtain $T_{\infty
}=6.869261369$.

A straightforward calculation shows that the approximate method outlined in
Sec.~\ref{sec:method} yields
\begin{equation}
T^{[1]}(\rho )=\frac{8\pi }{\sqrt{\rho (\pi ^{2}+4)+16}}  \label{eq:OD_T[1]}
\end{equation}
and $T_{\infty }^{[1]}\approx 6.75$ is a reasonable estimate of the exact
large--$\rho $ limit. These results were obtained by Ren and He\cite{RH09}.

For the small--$\rho $ series we have
\begin{equation}
T(\rho )=2\pi -\frac{8\rho }{3}+\ldots  \label{eq:DO_T_Taylor}
\end{equation}
and
\begin{equation}
T^{[1]}(\rho )=2\pi -\frac{\pi (\pi ^{2}+4)\rho }{16}+\ldots
\label{eq:DO_T[1]_Taylor}
\end{equation}

For the second approximation we have
\begin{eqnarray}
&&T^{6}\rho (\rho +1)^{2}(9\pi ^{2}-16)-8\pi ^{2}T^{4}(\rho +1)[\rho (45\pi
^{2}-16)+256]  \nonumber \\
&&+16\pi ^{4}T^{2}[\rho (369\pi ^{2}+1136)+5120]-294912\pi ^{6}=0
\label{eq:DO_T[2]}
\end{eqnarray}
and the expression for the large--$\rho $ limit
\begin{equation}
T_{\infty }^{6}(9\pi ^{2}-16)+8\pi ^{2}T_{\infty }^{4}(16-45\pi ^{2})+16\pi
^{4}T^{2}(369\pi ^{2}+1136)-294912\pi ^{6}=0  \label{eq:DO_T[2]_inf}
\end{equation}
The estimate $T_{\infty }^{[2]}\approx 6.866$ is more accurate than $%
T_{\infty }^{[1]}$.

The small--$\rho $ expansion is also more accurate
\begin{equation}
T^{[2]}(\rho )=2\pi -\frac{\pi (9\pi ^{2}+20)\rho }{128}+\ldots
\label{eq:DO_T[2]_Taylor}
\end{equation}
but neither $T^{[1]}$ nor $T^{[2]}$ gives more than the leading term exactly.

\begin{figure}[]
\begin{center}
\includegraphics[width=9cm]{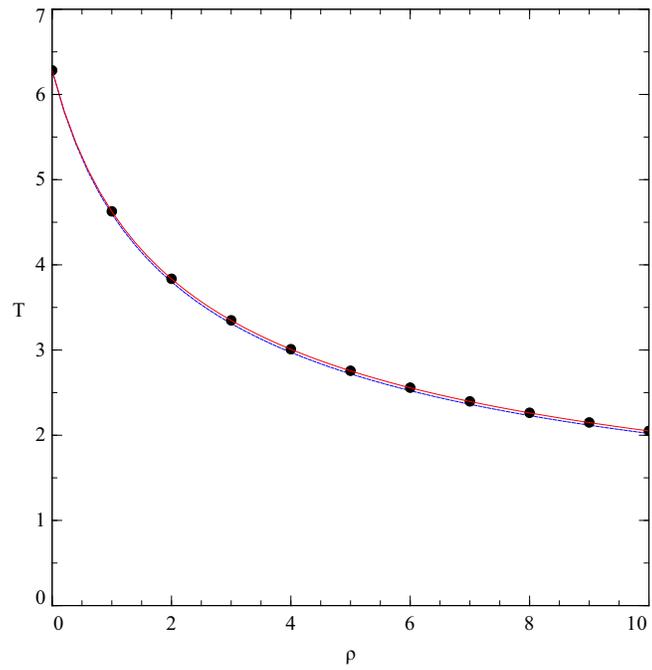}
\end{center}
\caption{Period for the discontinuous oscillator. Circles, dashed and solid
lines stand for exact, first and second approximations, respectively}
\label{fig:DO}
\end{figure}

Fig.~\ref{fig:DO} shows the exact period and the approximations of first and
second order for small and moderate values of $\rho $. Once again we
appreciate that the accuracy of the method increases with the order.

\section{Conclusions}

\label{sec:conclusions}

We have developed a straightforward systematic approach for the calculation
of the period of simple nonlinear oscillators. It is particularly useful if
we can integrate the equation (\ref{eq:u(t)[N]}) analytically which we
easily do when $f(u)$ is a polynomial or when a few terms of the Taylor
series about $u=0$ give a reasonable approximation.

Although present approach becomes increasingly complicated with the order,
we think that its equations are always considerably simpler than those
coming from the method of harmonic balance\cite{NM79}.

\end{document}